\documentstyle[11pt]{article}
             
\begin{document}
\newcommand{\beq}{\begin{equation}}
\newcommand{\eeq}{\end{equation}}
\newcommand{\bea}{\begin{eqnarray}}
\newcommand{\eea}{\end{eqnarray}}
\newcommand{\bal}{\begin{array}{ll}} \newcommand{\eal}{\end{array}}
\newcommand{\nn}{\nonumber}
\newcommand{\hl}{{\hat \lambda}}
\newcommand{\cm}{{\cal M}}
\def\Sup{\mathop{\rm\Sup}\nolimits}
\def\R {\rm R \kern -.35cm I \kern .19cm}
\def\C{ {\rm C \kern -.15cm \vrule width.5pt \kern .12cm}}
\def\Z{ {\rm Z \kern -.30cm \angle \kern .02cm}}
\def\1{ {\rm 1 \kern -.10cm I \kern .14cm}}
\def\N{ {\rm N \kern -.31cm I \kern .15cm}}

\begin{titlepage}
\title{\bf{Dynamical mass matrices from moduli fields \footnote{Based on
talks given at SUSY95, may 1995, Paris, EPS-HEP, july 1995, Brussels and
$5th$ Hellenic School and Workshop, september 1995, Corfu, to appear
in the Proceedings.}}}
\author{Emilian Dudas \\
        {\em Service de Physique Th\'eorique de Saclay}\\
        {\em 91191 Gif-sur-Yvette Cedex, France}
       }
\maketitle
\begin{abstract}
We review recent work on the structure of the fermion mass matrices in 
supergravity effective superstrings. They are generally given at low
 energy by
non-trivial functions of the gauge singlet moduli fields. Interesting
structures appear in particular if they are homogeneous functions of
zero degree in the moduli. In this case we find Yukawa  matrices very
similar to the ones obtained by imposing a $U(1)$ family symmetry 
to reproduce the observed hierarchy of masses and mixing angles.
The role of the $U(1)$ symmetry is played here by the modular symmetry.
The Flavor Changing Neutral Currents effects at the Planck scale coming
from the soft terms are identically zero.
A dynamical scenario is discussed which allow to
generate the observed hierarchies. 
\end{abstract}
\vskip 3cm
Saclay T96/012   \\hep-ph/9602231 \\ February 1996\\
\end{titlepage}
\section{Introduction}

One of the challenges of the Standard Model and its
extensions is the understanding of the fermion masses and mixings. These 
are usually arbitrary parameters and the  large hierarchy of masses
and mixing angles observed experimentally is still far to be understood. 

The solutions to this problem proposed to this date fall essentially
into two categories. One is a symmetry approach, first emphasized by
Froggatt and Nielsen \cite{FN}, which has been largely studied in the
literature. It postulates a new abelian horizontal gauge symmetry
spontaneously broken at a high energy scale $M_X$. The 3
families of quarks and leptons have different charges under the
corresponding $U(1)_X$ group so that only a small number of the
Standard Model Yukawa interactions be allowed by the symmetry $U(1)_X$.
All the others appear through non-renormalisable couplings to a
field whose vacuum expectation value $<\phi >$ breaks the horizontal
symmetry. In the effective theory below the scale of breaking, this
typically yields Yukawa couplings of the form 
\begin{equation}
\lambda_{ij} = \left( {<\phi > \over M_X} \right)^{n_{ij}},
\label{eq:X}
\end{equation}
where $n_{ij}$ depends on the $U(1)_X$ charges of the relevant fields.
If $\varepsilon \equiv <\!\phi\!>~/~M_X$ is a small parameter, the
hierarchy of masses and mixing angles is easily obtained by assigning
different charges for different fermions. 

A second approach, of dynamical origin, was recently proposed
\cite{N,Z,KPZ,BDu,BD}. The main idea is to treat Yukawa couplings as
dynamical variables to be fixed by the minimization of the vacuum energy
density. In this case, one can show that a large hierarchy can be
naturally obtained provided that the Yukawa couplings are subject to
constraints. Such a constraint could be obtained by an ad hoc imposition
of the absence of quadratic divergences in the vacuum energy
\cite{N,BDu} or as an approximate infrared evolution of the
renormalization group equations \cite{KPZ}. In ref. \cite{BD}, a
geometric origin for these constraints was proposed, related to the
properties of the moduli space in effective superstring theories. In
order to illustrate the idea, we give a simple example of a model
containing two moduli fields $T_1$, $T_2$ and two fermions with moduli
independent Yukawa couplings $\lambda_1$, $\lambda_2$. In this simple
model the low energy couplings at the Planck scale $M_P$ are simply
computed to be  \bea
\hat \lambda_1 \sim \left( {T_1 + T_1^+ \over T_2 + T_2^+} \right)^{3/4}
\lambda_1, \;\;\; \hat \lambda_2 \sim \left( {T_2 + T_2^+ \over T_1 +
T_1^+} \right)^{3/4} \lambda_2. \label{eq:Yuk}
\eea
Thus $\hat \lambda_1$ and $\hat \lambda_2$ are homogeneous functions of
zero degree in the moduli. 

The moduli fields correspond to flat directions in the
effective four-dimensional supergravity theory. If these flat
directions are exact, then the couplings in (\ref{eq:Yuk}) can be
regarded as dynamical variables to be determined by the low energy
physics (much in the spirit of the no-scale idea \cite{CFKN} used in the
dynamical determination of the gravitino mass \cite{ELNT}). It is easily
seen however that the product
\begin{equation}
\hat \lambda_1 \hat \lambda_2 \sim \lambda_1 \lambda_2 \label{eq:const}
\end{equation}
should be regarded as a constraint, because the moduli dependence has
disappeared in the right hand side of (\ref{eq:const}). Minimization of
the vacuum energy at a low energy scale with respect to the top
and bottom Yukawa couplings subject to a constraint of the type
(\ref{eq:const}) was studied in detail in \cite{BD}. It was shown there
that, qualitatively, the ratio of the two couplings behaves as
\begin{equation}
\left( {\lambda_b \over \lambda_t} \right) (\mu_0) \sim g^4 (M_P) {\mu
\over M_{SUSY}}, 
\end{equation}
wher $\mu_0 \sim 1 \; TeV$, $g(M_P)$ is the gauge coupling constant at
the Planck scale, $\mu$ is the usual supersymmetric mass parameter of
the MSSM and $M_{SUSY}$ is the typical mass splitting between
superpartners. For a large region of the parameter space of the MSSM,
one can thus obtain $\lambda_t / \lambda_b \sim 40-50$ and easily fit
the experimental masses with values of $\tan \beta$ of order 1.
Another interesting application of these dynamical ideas, related to
the flavor problem in supersymmetry \cite{DG} is not  discussed here.

The purpose of the present paper is twofold. First of all, we wish to
show that in effective superstrings of the orbifold type \cite{DHVW},
structures of the type (\ref{eq:X}) are naturally obtained. In this
approach, the small parameter $\varepsilon = <\phi> / M_X$ of the
$U(1)_X$ horizontal symmetry is given here by $\varepsilon = (T_1 +
T_1^+) /  (T_2 + T_2^+)$ in the case of two moduli. 

The second goal is to show that one of the cases leading to these
Froggatt-Nielsen structures\footnote{ From now on, we will call
Froggatt-Nielsen structures mass matrices for which the order of
magnitude of all entries are rational powers of a small common quantity
$\varepsilon$.} corresponds precisely to Yukawa couplings being
homogeneous functions of the moduli. We can then apply the
above-mentionned dynamical mechanism and determine by minimization the
whole structure of the fermion matrices. The hierarchy translates into
different vacuum expectation values of the moduli fields and different
modular weights of the fermions with respect to these moduli.

Section 2 presents all the cases corresponding to Froggatt-Nielsen
structures in orbifold-like effective models. In some instances, they
appear when some -- but not all -- moduli fields are fixed to their
self-dual values $T_\alpha = 1$. An appealing situation is the case where the
theory possesses a ``diagonal'' modular symmetry; then the Yukawa
couplings are homogeneous functions of zero degree in the moduli and
Froggatt-Nielsen structures appear even if all $T_\alpha$ are different from
1. Remarkably enough, there are no contributions at all to the FCNC effects
at the Planck scale coming from the soft terms. This is to be contrasted
to the $U(1)_X$ horizontal symmetry approach, where departures from
universality are difficult to avoid. Moreover, the predictions we find
for the soft terms are a weaker form of the conditions obtained by imposing 
the 2-loop ultraviolet finiteness of the theory, recently discussed in
 the literature. 

Section 3 analyses, in analogy with the $U(1)_X$ approach, the
relation between the mass matrices and the one-loop modular anomalies.
It is shown that if there are no string threshold corrections in the
gauge coupling constants the anomalies can be eliminated only by the
Green-Schwarz mechanism \cite{GS} which uses the Kalb-Ramond
antisymmetric tensor field present in superstring theories. In the
case relevant for the dynamical approach, the modular anomalies can be
cancelled by this mechanism only if there exists at least two moduli 
with modular anomalies cancelled by the Green-Schwarz mechanism.
If the threshold corrections are present, they can account for a part
of the modular anomalies and can provide a correct gauge coupling
unification scenario.

Section 4 deals with the dynamical determination of the mass matrices
at low energy along the lines of ref. \cite{BD}. Two additional
constraints on the modular weights are needed for the mechanism to be
effective.

Some conclusions are presented at the end, together with open
questions that remain to be investigated. 
\section {Low-energy mass matrices.} \vskip .5cm
The low energy limit of the superstring models relevant for the phenomenology is the $N=1$
supergravity described by the K\"ahler function $K$, the superpotential $W$ and the gauge
kinetic function $f$ \cite{CFGP} . The generic fields present in the zero-mass string spectrum
contain an universal dilaton-like field $S$, moduli fields generically denoted by $T_\alpha$
(which can contain the radii-type moduli $T_\alpha$ and the complex
structure moduli  $U_\beta$) and
some matter chiral fields $\phi^i$, containing the standard model particles. The K\"ahler potential
and the superpotential read
\bea
K &=& K_0 + \sum_i \prod_{\alpha} 
t^{n^{(\alpha)}_i}_{\alpha} |\phi^i |^2 + \cdots , \nn \\
K_0 &=& {\hat K}_0(T_\alpha , T^+_\alpha ) - \ln (S+S^+) ,
\label{eq:ar1}\\ W &=& {1 \over 3} \lambda_{ijk} \phi^i \phi^j \phi^k +
\cdots , \nn \eea
where the dots stand for higher-order terms in the fields $\phi^i$. In \ (\ref{eq:ar1}), $t_\alpha =
Re T_\alpha$ are the real parts of the moduli and $n_i^{(\alpha )}$ are called the
modular weights of the fields $\phi^i$ with respect to the modulus
$T_\alpha$ . The $\lambda_{ijk}$ are the  Yukawa couplings which
may depend nonperturbatively on $S$ and $T_\alpha$. We define the
diagonal modular weight of the field $\phi^i$ as
$n_i = \sum_\alpha n_i^{(\alpha)}$ . 
An important role in the following discussion will be played by the target-space modular
symmetries\ $SL(2,\Z)$\ associated with the moduli fields $T_\alpha$ , acting as
\bea
T_\alpha \rightarrow {a_\alpha T_\alpha - i b_\alpha \over i c_\alpha 
T_\alpha + d_\alpha} ,\; a_\alpha d_\alpha - b_\alpha c_\alpha = 1 ,\;
a_\alpha \cdots d_\alpha \in \Z \;\; . \label{eq:ar3} 
\eea
In effective string theories of the orbifold type \cite{DHVW}, the matter
fields \ $\phi^i$ transform under
(\ref{eq:ar3}) as \bea
\phi^i \rightarrow  (i c_\alpha T_\alpha + d_\alpha )^{n_i^{(\alpha)}}
\phi^i  \label{eq:ar4}
\eea
in order for the K\"ahler metric $K_i^j = \partial^2 K / \partial
\phi^i \partial \bar \phi_j$ to be invariant.

A typical example is a model with $n$ moduli fields and
K\"ahler potential
\bea
K_0 = -  \sum_{\alpha = 1}^n p_{\alpha} \ln (T_\alpha + T_\alpha^+ ) - 
\ln (S+S^+) . \label{eq:ar6} \eea
Under (\ref{eq:ar3}), it transforms as
\bea
K \rightarrow K +  p_{\alpha} \ln | i c_\alpha T_\alpha + d_\alpha |^2
\label{eq:ar7} \eea

Associating a modular weight $n^{(\alpha )}_{ijk}$ with the trilinear
couplings in eq.(\ref{eq:ar1}), the transformation  of $W$
gives 
\bea 
n^{(\alpha )}_i + n_j^{(\alpha )} + n_k^{(\alpha)} +
n^{(\alpha )}_{ijk} = - p_{\alpha} . \label{eq:ar8} 
\eea
Taking the sum of all such relations for the moduli fields, we find
\bea 
n_i + n_j + n_k + n_{ijk} = - p . \label{eq:ar9}
\eea
Eq.(\ref{eq:ar9}) is a weaker form of eqs.(\ref{eq:ar8}),
expressing the invariance of the theory under the diagonal modular
transformations with respect to all moduli~: 
\bea
\phi^i \rightarrow \prod_\alpha (ic_\alpha T_\alpha + d_\alpha
)^{n_i^{(\alpha)}} \phi^i. \label{eq:ar10'}
\eea
The difference between the
individual modular transformations and the less restrictive diagonal one
will be essential in the following.

The low energy spontaneously broken theory contains the canonically 
normalized field $\hat \phi^i$ defined by $\phi^i = (K^{-1/2})^i_j \
\hat{\phi^j}$ and the Yukawas $\hat \lambda_{ijk}$ which give the
physical masses. The matching condition at the Planck scale $M_p$
relating the low energy and the original Yukawa couplings is \bea
\hat \lambda_{ijk} = e^{K_0 \over 2} (K^{-1/2})_i^{i'} (K^{-1/2})_j^{j'} (K^{-1/2})_k^{k'}
\lambda_{i'j'k'} \ . \label{eq:ar10}
\eea
From eq.(\ref{eq:ar10}) we see that the $\hat \lambda_{ijk}$ are
functions of the moduli through the K\"ahler potential $K$ and
eventually the $\lambda_{i'j'k'}$.
The possible dependence of $\lambda_{ijk}$ on the moduli fields in
connection with the fermion mass matrices was analyzed in detail in
the literature \cite{CGM}. In most of the following considerations we
consider only the case $\lambda_{ijk} = cst + O(e^{-T})$, in the limit
where the moduli dependence can be neglected (for example, large
compactification radius limit). The Yukawa matrices in fermionic
string constructions were studied in detail, too (see, e.g. \cite{FARAGGI}).
 
Our goal is to analyze the general structure of the mass matrices for 
the quarks and leptons as a
function of the moduli fields. They are described by the superpotential
$\hat W$ of the Minimal Supersymmetric Standard Model (MSSM) which we
take to be the minimal model obtained in the low-energy limit of the
superstring models, plus eventually some extra matter singlet under the
Standard Model gauge group. $\hat W$ contains the Yukawa interactions
\bea 
\hat W \supset \hat \lambda_{ij}^U \hat Q^i \hat U^c_j \hat H_2 +
\hat \lambda^D_{ij} \hat Q^i \hat D^c_j \hat H_1 + \hat \lambda^L_{ij}
\hat L^i \hat E^c_j \hat H_1 , \label{eq:ar11} 
\eea
where $H_1$ and $H_2$ are the two Higgs doublets of MSSM , $Q^i$, $L^i$ are the $SU(2)$ quark
and lepton doublets and $U_j$, $D_j$ , $E_j$ are the right-handed
$SU(2)$ singlets.

Consider the case of two moduli $T_1$ and $T_2$. Using eqs.
(\ref{eq:ar1}),
(\ref{eq:ar10}) and (\ref{eq:ar11}) , a $U$-quark coupling reads
\bea
\hat \lambda^U_{ij} = e^{K_0 \over 2} \ t_2^{- {n_{Q_i} + n_{U_j} + n_{H_2} \over 2}} {\big ({t_1
\over t_2}\big ) ^{- {n_{Q_i}^{(1)} + n_{U_j}^{(1)} + n_{H_2}^{(1)} 
\over 2}}} \lambda^U_{ij}\ \label{eq:ar12}
\eea
or equivalently,
\bea
\hat \lambda^U_{ij} = e^{K_0 \over 2} \ t_1^{- {n_{Q_i} + n_{U_j} + n_{H_2} \over 2}} {\big ({t_2
\over t_1}\big ) ^{- {n_{Q_i}^{(2)} + n_{U_j}^{(2)} + n_{H_2}^{(2)} \over 2}}} \lambda^U_{ij}
\ . \label{eq:ar13}
\eea
{\em Suppose that one of the two moduli-dependent factors in
(\ref{eq:ar12}) (or equivalently  (\ref{eq:ar13})) happens to be family
blind. Then the structure obtained for the Yukawa matrix turns out to be
very similar to the one that would be derived from an horizontal $U(1)$
symmetry  of the Froggatt-Nielsen type \cite{FN}. Modular weights play
the role of the $U(1)$ charges.} Such a situation may arise in the
following three cases of interest:

i) $t_1 = t_2 = t \not= 1$.

Then
\bea
\hat \lambda^U_{ij} =   t^{- {n_{Q_i} + n_{U_j} + n_{H_2} + p \over 2}} 
\ \lambda^U_{ij}
\ , \label{eq:ar130}
\eea
where $n_{Q_i}$, etc are the diagonal modular weights. 
For $t << 1$ this could produce hierarchical Yukawa couplings. This case
is disfavoured in the case where the relation (\ref{eq:ar9}) holds with
$n_{ijk} = 0$.

ii) $t_2 = 1 , {t_1 \over t_2} = \varepsilon <<  1 $ or vice versa $t_1 \leftrightarrow t_2$.
Then using eq.(\ref{eq:ar12}) we get
\bea
\hat \lambda^U_{ij} \sim 
\varepsilon^{-{ n_{Q_i}^{(1)} + n_{U_j}^{(1)} + n_{H_2}^{(1)} \over 2}}
\lambda^U _{ij} , \label{eq:ar14}
\eea
where we dropped the universal $e^{K_0 \over 2}$ factor, irrelevant
here. Hierarchical structures are obtained if the dynamics imposes
$\varepsilon = {t_1 \over t_2}$ small (typically of the order of the
Cabibbo angle to some power). Remark that the relevant modular weights
correspond to  the modulus whose ground state falls away from the
self-dual points, $t_i \not= 1$ (for an example of such
a situation, see Ref.\cite{MacRoss}). 

iii) one has the condition
\bea
n_{Q_i} + n_{U_j} + n_{H_2} = \ \mbox{independent of }\  i\  \mbox{and} 
\ j . \label{eq:ar15}
\eea
and ${t_1 \over t_2} = \varepsilon <<  1 $ or vice-versa.
This obviously implies that $n_{Q_i} = n_Q$ and $n_{U_i} = n_U$
for any $i = 1, 2, 3$.

For example, in the case when the diagonal modular symmetry holds, the
constant (\ref{eq:ar15}) is equal to $-p$ and the Yukawa couplings can be
written as 
\bea  
\hat \lambda_{ij}^U = e^{K_0 \over 2} {t_2}^{p \over 2}
\left( {t_1 \over t_2} \right)^{-{ n_{Q_i}^{(1)} + n_{U_j}^{(1)} +
n_{H_2}^{(1)} + {3 \over 2} \over 2}}  \lambda_{ij}^U \; = \; 
e^{K_0 \over 2} {t_1}^{p \over 2} \left( {t_2 \over t_1} \right)^{-{
n_{Q_i}^{(2)} + n_{U_j}^{(2)} + n_{H_2}^{(2)} + {3 \over 2} \over 2}} 
\lambda_{ij}^U \label{eq:ar15a}
\eea
or
\bea
\hat \lambda_{ij}^U = e^{K_0 \over 2} {(t_1 t_2)}^{p \over 4} \left( {t_1 \over t_2}
\right)^{-{ n_{Q_i}^{(12)} + n_{U_j}^{(12)} + n_{H_2}^{(12)} \over
4}}  \lambda_{ij}^U \label{eq:ar15b}
\eea
where $n_{Q_i}^{(12)} = n_{Q_i}^{(1)} - n_{Q_i}^{(2)}$, etc. The last
form (\ref{eq:ar15b}) is particularly useful in that it relates the
Froggatt-Nielsen-like structures with the asymmetry between the modular
weights corresponding to the two moduli fields. Notice that the structures
survives in the large radius limit $T_{\alpha} \rightarrow \infty$,
because only ratios of moduli fields appear in (\ref{eq:ar15b}).

One way to get the condition (\ref{eq:ar15}) is to search for models
where the $\hat \lambda_{ijk}$ are \underline{homogeneous functions} of
the moduli $T_\alpha$, i.e. $\sum_\alpha t_\alpha \partial \hat
\lambda_{ijk} / \partial t_\alpha = 0$. In this case, using the
relation $\sum_\alpha t_\alpha \partial  K^j_i / \partial t_\alpha
= n_i K^j_i$ and the matching condition (\ref{eq:ar10}), we arrive at an
equation for the original  couplings $\lambda_{ijk}$ 
\bea
\big ( {1 \over2} t_\alpha K^\alpha - {n_i + n_j + n_k \over 2} + t_\alpha {\partial \over \partial
t_\alpha} \big ) \lambda_{ijk} = 0 . \label{eq:ar16}
\eea
If $t_\alpha K^\alpha = - p$ and the $\lambda_{ijk}$ are pure
numbers we recover  eq.(\ref{eq:ar9}). In such a case ($n_{ijk} =
0$), the relation (\ref{eq:ar16}) can be derived from assuming the
diagonal modular symmetry discussed above in (\ref{eq:ar10'}). This
approach was used in \cite{BD} in a dynamical approach to the fermion
mass problem proposed in \cite{N}, \cite{Z}  and studied in \cite{KPZ}
and \cite{BDu}. We will return to it in section $4$.

The experimental data on the fermion and the mixing angles can be summarized as follows.
Defining $\lambda = sin \theta_c \sim 0.22$ where $\theta_c$ is the Cabibbo angle, the mass
ratios and the Kobayashi-Maskawa matrix elements at a high scale $M_X
\sim M_P$ have the values \bea
&{m_u \over m_t} \sim \lambda^7 \; {\rm to} \; \lambda^8 \ , 
\  {m_c \over m_t} \sim \lambda^4 \ , 
\  {m_d \over m_b} \sim \lambda^4\  , 
\  {m_s \over m_b} \sim \lambda^2\ ,\nn \\ 
&{m_e \over m_\tau } \sim \lambda^4 \ , 
\ {m_\mu \over m_\tau } \sim \lambda^2 \ , 
\ | V_{us} | \sim \lambda \ ,
\ | V_{cb} | \sim \lambda^2 \ , 
\ | V_{ub} | \sim \lambda^3 \; {\rm to} \;  \lambda^4
. \label{eq:ar19} 
\eea 
Taking as a small parameter $\varepsilon =
\lambda^2 \sim {1 \over 20}$, these values are perfectly
acommodated by the following modular weight assignement 
\bea
n_{Q_3}^{(1)} - n_{Q_1}^{(1)} = 3 \ , \ n_{Q_3}^{(1)} - n_{Q_2}^{(1)} =
2 \nn \\ n_s^{(1)} - n_b^{(1)} = 0 \ , \ n_b^{(1)} - n_d^{(1)} = 1 
\label{eq:ar20} \\ n_t^{(1)} - n_c^{(1)} = 2 \ , \ n_t^{(1)} - n_u^{(1)}
= 4 \nn  
\eea 
corresponding to the mass matrices 
\bea
\hat \lambda_U = {\lambda^{-x}} \left ( \begin{array}{cc} 
\lambda^7 \ \lambda^5 \   \lambda^3    \\ 
\lambda^6 \   \lambda^4  \  \lambda^2    \\
\lambda^4 \  \lambda^2 \ \ 1 
\end{array}  \right ) , \;\;
\hat \lambda_D = {\lambda^y} \left ( \begin{array}{cc} 
\lambda^4 \ \lambda^3 \ \lambda^3   \\ 
\lambda^3 \  \lambda^2 \ \lambda^2   \\ 
\lambda \ \ 1 \ \ 1  
\end{array}  \right ) . \label{eq:ar21}
\eea
In (\ref{eq:ar21}) $x \ge 0$ (the top coupling should be at least of 
order one at a high scale) and $y \ge 0$ (the bottom coupling should
correspondingly be smaller or equal to one). Remark that the negative
power of $\lambda$ in $\hat \lambda_U$ is impossible to obtain in a
horizontal symmetry approach because of the analyticity of the
superpotential. 

An interesting aspect of the case iii) discussed above should be
stressed which concerns the sfermion masses $(M^2_0)_{i \bar \jmath}$.
Non-diagonal sfermion mass matrices give potentially
dangerous contributions to flavor changing neutral current processes
like $b\rightarrow s \gamma$ or $\mu \rightarrow e \gamma$ \cite{GM}. The
general expression in supergravity is  \bea (M^2_0)_{i \bar \jmath} =
(G_{i \bar \jmath} - G_\alpha
R^\alpha_{{\bar \jmath}i \beta } G^\beta ) m_{3/2}^2 , \label{eq:ar22} 
\eea
where $G = K + l n | W |^2$ and $ G_{i \bar \jmath} = {\partial^2 G 
\over \partial \phi^i \partial \phi^+_{\bar \jmath}} $
is the metric on the K\"ahler space.
The indices $\alpha , \beta $ correspond to moduli fields which contribute to supersymmetry
breaking $< G_\alpha > \not= 0$ and we assume $< G_\alpha G^\alpha > +
< G_S G^S > = 3 $ \cite{BIM}.  $R^\alpha_{{\bar
\jmath}i\beta}$ is the Riemann tensor of the K\"ahler space and $m_{3/2}$ the
 gravitino mass. 
The trilinear soft breaking terms $A_{ijk}$ associated with the Yukawa
couplings $W_{ijk}$ are given by \cite{FKZ}
\bea
A_{ijk} = \left[ ( 3 + G^{\alpha} D_{\alpha} + G^S D_S) {W_{ijk} \over W}
\right] m_{3/2}^2 \ . \label{eq:ar220}
\eea
 
If the superpotential does not depend on the moduli fields,
$< G_\alpha G^\alpha > = p$. In this case, a straightforward supergravity
computation for the soft terms gives (M is the universal gaugino mass)
\bea
{\tilde m}_Q^2 = m_{3/2} ^2 (1 + n_Q) \1 
,\nn \\
A_{ij}^U = -M + (n_Q + n_U + n_{H_2} + p) m_{3/2} , 
\nn \\ 
M^2 = (3-p) m_{3/2}^2 , 
\label{eq:ar24}
\eea
where ${\tilde m}_Q^2$ is the left-left squark
squared mass and similar relations with obvious replacements hold for 
all the other fermions. In
(\ref{eq:ar24}) $n_Q$ and $n_U$  are the diagonal modular weights
which in the case iii) discussed above are the same for the three
generations. Using the modular invariance conditions (\ref{eq:ar9}) with
$n_{ijk}=0$, we find the soft terms predictions at the Planck scale 
\cite{BIND}
\bea
A_{ij}^U = A_{ij}^D = A_{ij}^L = M 
,\nn \\
{\tilde m}_Q^2 + {\tilde m}_U^2 + {\tilde m}_{H_2}^2 = M^2  , 
\nn \\ 
{\tilde m}_Q^2 + {\tilde m}_D^2 + {\tilde m}_{H_1}^2 = M^2  , 
\nn \\ 
{\tilde m}_L^2 + {\tilde m}_E^2 + {\tilde m}_{H_1}^2 = M^2  . 
\label{eq:ar240}
\eea

Notice that the relations (\ref{eq:ar240}) are a weaker form of the ultraviolet
finiteness
conditions for the soft terms in finite theories, widely studied in the last 
years \cite{JMY}, \cite{EKT} (they can be also obtained as infrared fixed
point predictions of an underlying theory \cite{FJJ}). This surprising
connection between finiteness and our predictions can be technically
traced back to the modular invariance of the theory and the assumption that
supersymmetry breaking is saturated by the moduli and dilaton fields.

The important aspect of (\ref{eq:ar24}) is that the sfermion
soft-breaking mass matrix and the trilinear soft matrices are proportional
to the identity. Going to the
basis where the quark mass matrices are diagonal, we see that the soft
squark masses are still proportional to the unit matrix. Consequently,
there are no flavor changing neutral currents  induced at the
supergravity level. It is well known that usually models of fermion masses
based on horizontal symmetries have problems with FCNC processes due to the
soft terms \cite{NS}, \cite{DPSA}. 
As far as I know, the models discussed here are the only known examples
 with fermion
mass matrices of Froggatt-Nielsen type \cite{FN} and exact flavor independence
of soft terms at the Planck scale.
Even if we know by now \cite{BIM} that SUGRA -
induced flavor changing neutral currents are not as severe as thought
several years ago, it is still worth emphasizing the virtue of the case
iii), corresponding to considering Yukawa as being homogeneous functions
of the moduli.  
\vskip 1cm
\section {Modular anomalies and moduli mass
textures.}  \vskip .5cm
 
In the context of horizontal abelian symmetries used to explain fermion
mass hierarchies, an interesting connection has been established
\cite{IR,BR} between anomalies associated with such symmetries and mass
hierarchies as given in (\ref{eq:ar19}). 

Gauge anomalies in usual field theories must be absent in
order to define a consistent quantum theory. This requirement  imposes
non-trivial constraints on the particle spectrum in chiral theories.
Applied to the ten-dimensional superstrings, this led to the famous 
Green-Schwarz anomaly cancellation mechanism \cite{GS}. This mechanism
 has a counterpart
in 4 dimensions which allows to fix the value of $\sin ^2 \theta_W$ at
the string scale without advocating a grand unified symmetry \cite{IB}.
It was shown in \cite{BR} and generalized in \cite{DPS,Nir} that, using
this mechanism, it is possible to infer from the observed hierarchies
(\ref{eq:ar19}) in the mass matrices the standard value of $3/8$ for
$\sin ^2 \theta_W$.

Effective string models also have another type of anomalies,
named $\sigma$-model anomalies \cite{MN}. They appear in triangle
 diagrams with two gauge bosons and one modulus.  
 We will show
in this section that the cancellation of these anomalies plays a role
very similar to the one of mixed gauge anomalies in the
abelian horizontal $U(1)_X$ symmetry approach.    

The cases of interest to be analyzed in this paper are orbifold
compactifications. Consider the diagonal K\"ahler moduli for which 
$\hat K_0 (T_\beta, T^+_\beta) = - \ln (T_\beta + T^+_\beta)$ (and
possibly the complex structure moduli). 
The gauge group is
$G = \prod_a G_a$ and there are matter fields in different
representations $R_a$ of $G_a$. The anomalous triangle diagrams give a
non-local contribution to the one-loop effective lagrangian \cite{DFKZ}
which reads  
\beq 
{\cal L}_{nl } = {1 \over 8} {1 \over 16 \pi^2} \sum_a 
\int d^4 \theta  (W^\alpha W_\alpha)_a {{\cal D}^2 \over \Box }
\sum_\beta b_a'^{(\beta)} \ln (T_\beta + T^+_\beta )
 + h.c. \label{eq:ar26} 
\eeq

In eq.(\ref{eq:ar26}) where superfield notations are used, $W^\alpha$ is
the Yang-Mills field strength superfield and $b_a'^{(\beta)}$ are the
 anomaly coefficients. 

The change of ${\cal L}_{nl }$ under the modular
transformations (\ref{eq:ar3}) is given by the local expression
\bea
\delta {\cal L}_{nl } = {1 \over 2} {1 \over 16 \pi^2 }  \sum_a
\int d^2 \theta (W^\alpha W_\alpha)_a \sum_\beta  b_a'^{(\beta)}
\ln (i c_\beta T_\beta + d_\beta ) + h.c. \label{eq:ar28}  
\eea

There are two ways of compensating this anomaly. The first, which  is particularly interesting
in our case is reminiscent of the Green-Schwarz mechanism. It
 requires the
non-invariance of the dilaton field under the modular transformations
$ S \rightarrow S - {1 \over 8\pi^2} \sum_\beta \delta^{(\beta)}_{GS} 
\ln (i c_\beta T_\beta + d_\beta ) . $
The factor $\delta_{GS}^{(\beta)}$ is the gauge group independent Green-Schwarz coefficient
and induces a mixing between the dilaton $S$ and the moduli fields
$T_\beta$. This mechanism can completely cancel the anomalies  only if
the anomaly coefficients $b_a'^{(\beta)}$ satisfy the equalities
$\delta_{GS}^{(\beta)} = {b_a '^{(\beta )} \over k_a} = {b_b'^{(\beta)} \over k_b} = \cdots $
for all the group factors of the gauge group $G = \prod_a G_a$.

A second mechanism for the cancellation of the term (\ref{eq:ar28}) uses the one-loop threshold
corrections to the gauge coupling constants, which can be different for different gauge group
factors.  If the modular symmetry group
is  $(SL(2,\Z))^3$, the one-loop running gauge coupling constants 
at a scale $\mu$ reads
\bea
{1 \over g^2_a (\mu)} = {k_a \over g^2_s} + {b_a \over 16 \pi^2} ln {M^2_s \over \mu^2} - {1 \over 16 \pi^2} \sum_{\alpha =1}^3 (b'^{(\alpha)}_a - k_a {\delta
^{(\alpha)}_{GS}}) \ln \left [ (T_{\alpha} + T^+_{\alpha}) 
|\eta (T_{\alpha})|^4 \right ]  \label{eq:ar131}
\eea
In (\ref{eq:ar131}) $g_s$ is the string coupling constant, $M_s$ is the string
scale and $b_a$ are the RG $\beta$-function coefficients ($a = 1,2,3$)  for
$U(1)_Y$, $SU(2)_L$ and $SU(3)$ respectively. The
Dedekind function is defined by \ $\eta (T) = exp { (- \pi T / 12)
} \prod_{n=1}^{ \infty} [1 - exp {(2 \pi n T)}]$, which transforms under
(\ref{eq:ar3}) as  $\eta (T_{\alpha}) \rightarrow \eta (T_{\alpha}) (i
c_{\alpha} T_{\alpha} +  d_{\alpha})^{1 \over 2}$.
The unification scale $M_U$ is  computed to be
\bea
M_U = M_s \prod_{\alpha =1}^3  \left [ (T_{\alpha} + T^+_{\alpha}) 
| \eta (T_{\alpha})|^4 \right ]^{b'^{(\alpha)}_b k_a - b'^{(\alpha)}_a k_b \over 2
(b_a k_b - b_b k_a)} \ . \label{eq:ar132}
\eea
with $a \not= b \in \{ 1,2,3\}$.
In the simplest approximation of neglecting all the threshold corrections, a
one-loop RG analysis for $g_3$ and $sin^2 \theta_W$ gives a good agreement with
the experimental data if $M_U \simeq M_s / 50$ \cite{CEFNZ}.

Consider now a minimal orbifold model with the particle content of 
the \linebreak MSSM (respectively $Q_i,U_i,D_i,L_i$, $i$ being a family
index, and the two Higgs supermultiplets $H_1$ and $H_2$), plus possibly
extra Standard Model singlet fields. The mixed K\"ahler $SU(3) \times
SU(2) \times U(1)_Y$ triangle anomalies are described by the
coefficients \cite{IL} 
\bea 
&b_1 '^{(\beta )} &= 11 + \sum^3_{i=1} ({1 \over 3} n^{(\beta )}_{Q_i } + {8 \over 3} n^{(\beta
)}_{U_i} + {2 \over 3} n^{(\beta )}_{D_i} + n^{(\beta )}_{L_i} + 2n^{(\beta )}_{E_i} ) + n^{(\beta
)}_{H_1} +  n^{(\beta )}_{H_2} , \nn   \\
&b_2 '^{(\beta )} &= 5 + \sum^3_{i=1} (3n^{(\beta )}_{Q_i } + n^{(\beta )}_{L_i}) + n^{(\beta
)}_{H_1} + n^{(\beta )}_{H_2} , \label{eq:ar31} \\
&b_3 '^{(\beta )} &= 3 + \sum^3_{i=1}
(2n^{(\beta )}_{Q_i } + n^{(\beta )}_{U_i} + n^{(\beta )}_{D_i} ). \nn 
\eea
Apart from the modular weight independent piece, these coefficients
are identical to the ones encountered for the
mixed $U(1)_X - G_a$ gauge group anomalies in the abelian horizontal
$U(1)_X$ gauge symmetry approach. Again, the role of the $U(1)_X$ charges
is played here by the modular weights of the different fields.
Consequently we will closely follow  the analysis performed in  \cite{IR}
, \cite{BR} and \cite{DPS}. 

We place ourselves in the case  (iii) of the preceding section. Starting from 
the relation (\ref{eq:ar15b}), we
obtain: 
\bea
({\rm Det} \hl_U ) ({\rm Det} \hl_L)^3 ({\rm Det} \hl_D)^{-2} &\sim &
\varepsilon^{-{3 \over 8} (b_1'^{(12)} + b_2'^{(12)} - 2b_3'^{(12)})},
\nn \\  
{ Det \hl_L \over Det \hl_D } &\sim & \varepsilon^{-{1 \over 8}
[b_1'^{(12)} + b_2'^{(12)}  - {8 \over 3} b_3'^{(12)} - 2
(n_{H_1}^{(12)} + n_{H_2}^{(12)} )] } \ .\label{eq:n47}   
\eea
The first of eqs. (\ref{eq:n47}) is very useful to discuss anomaly
cancellation conditions. Taking as an example $\varepsilon \sim
\lambda^m$, it requires
\beq
b_1'^{(1)} + b_2'^{(1)} - 2 b_3'^{(1)} = b_1'^{(2)} + b_2'^{(2)} - 2
b_3'^{(2)} - {48 \over m}. \label{eq:n48}
\eeq
As shown in \cite{BR} for the case of an horizontal symmetry, the second
of eqs. (\ref{eq:n47}) has the following interesting solution, which
automatically gives the value $3/8$ for $\sin^2 \theta_W$ at
unification:
\bea
 n_{H_1}^{(1)} + n_{H_2}^{(1)} &=& n_{H_1}^{(2)} + n_{H_2}^{(2)}, \nn \\
b_1'^{(1)} + b_2'^{(1)} - {8 \over 3} b_3'^{(1)} &=& b_1'^{(2)} +
b_2'^{(2)} - {8 \over 3} b_3'^{(2)} . \label{eq:n49}
\eea

Moreover, using the conditions (\ref{eq:ar9}) in the case $n_{ijk} = 0$
and the expressions (\ref{eq:ar31}), we obtain
\bea
b'_1 + b'_2 - 2 b'_3 &=& 8, \nn  \\
b'_1 + b'_2 - {8 \over 3} b'_3 &=& 2 (8 + n_{H_1} + n_{H_2}),
\label{eq:n50}
\eea
where $b'_1 = b_1'^{(1)} + b_1'^{(2)}$, etc. Eqs. (\ref{eq:n48}) and
(\ref{eq:n50}) clearly express the fact that the theory has one-loop
modular anomalies.

An analysis of all the possibilities for the anomalies related to the
two moduli leads to the conclusion that, without threshold corrections,
the mixed case with zero anomalies for one modulus and Green-Schwarz
mechanism for the other modulus is physically uninteresting (it
requires $\varepsilon \sim \lambda^{\pm 6}$). In the case of anomalies
cancelled by the Green-Schwarz mechanism for both moduli, we obtain 
$n_{H_1}^{(1)} + n_{H_2}^{(1)} = n_{H_1}^{(2)} + n_{H_2}^{(2)}=-4$
and $b'^{(i)} = 6 ( 1 \pm 6/m)$ for $i=2, 1$. The only other allowed
case is when threshold corrections are present for both moduli. In this
case, we obtain $n_{H_1} + n_{H_2} + 8 = -{60 \lambda^m \over \pi} \ln
(M_S^2 / M_U^2)$. A realistic value for $M_U$ requires $m \ge 2$ and is
obtained for example for $m=2$, $n_{H_1} + n_{H_2}= -14$. Let us note
that $\sin^2 \theta_W$ can still be found equal to $3/8$ at unification
scale, irrespective of the choices made in order to obtain the desired
value for $M_U$.

\vskip 1cm

\section {Dynamical determination of couplings.}
\vskip .5cm
The duality symmetries imply the existence of flat directions in the
corresponding moduli fields. If they are respected to all orders in
the supergravity interactions, then the only way to lift them is 
by breaking supersymmetry. Given the scale expected for this breaking,
one may expect the low energy sector to play an important role in the
determination of the moduli ground state. Under these conditions, the
low energy minimization with respect to the moduli fields is
presumably equivalent to the minimization with respect to the Yukawa
couplings, through their non-trivial dependence on the moduli. This was
the attitude taken in Refs. \cite{KPZ,BDu,BD} to dynamically determine
the top/bottom Yukawa couplings. A very important point in this program
is the existence of constraints between Yukawas , of a type which
is typical of the approach based on moduli dynamics. It was shown in
Ref. \cite{BD} that this can be enforced if the Yukawa couplings are
homogeneous functions of the moduli. In what follows, we will
therefore place ourselves in the case iii) of section $2$ and analyze
how the two approaches can be merged, leading to a dynamical
determination of the fermion mass hierarchies and mixing angles. 

We start by reviewing the results of Ref. \cite{BD}. To
compute the vacuum energy at the low-energy scale $\mu_0 \sim M_{susy}$
we proceed in the usual way. Using boundary values compatible with the
constraints at the Planck scale $M_P$ (identified here with the
unification scale), we evolve the running parameters down to the scale
$\mu_0$ using the RG equations and adopt the effective potential
approach \cite{CW}. The one-loop effective potential has two pieces  \bea
V_1 (\mu_0) = V_0 (\mu_0) + \Delta V_1 (\mu_0) \ , \label{eq:ar101}
\eea 
where $V_0 (\mu_0)$ is the renormalization group improved tree-level
potential and $\Delta V_1 (\mu_0)$ summarizes the quantum corrections
given by the formula
\bea
\Delta V_1 (\mu_0) = {(1 / 64 \pi^2)} \ Str \cm^4 \ (\ln {\cm^2 \over
\mu_0^2} - {3 \over 2}) \ . \label{eq:ar102}
\eea
In (\ref{eq:ar102}) $\cm$ is the field-dependent mass matrix, 
$Str \cm^n = \sum_J (-1)^{2J} (2J+1) \ Tr M^n_J$ is the ponderated trace
of the mass matrix for particles of spin $J$  and all the
parameters are computed at the scale $\mu_0$. The vacuum state is
determined by the equation $\partial V_1 / \partial \phi_i = 0$, where
$\phi_i$ denotes collectively all the fields of the theory. The vacuum
energy is simply the value of the effective potential computed at the
minimum.\footnote{ In a first approximation, if the moduli masses are
larger than the average superpartner mass ${\tilde m}$, the factor
$\ln \cm^2 / \mu_0^2$ in (\ref{eq:ar102}) can be replaced by $\ln
{\tilde m}^2 / \mu_0^2 < 0$. \cite{BDu,BD}}
 
As expected there is no Yukawa coupling dependence at the tree level.
At the one-loop level it appears through
\bea
 \bal { 1 \over 3} \ Str \cm^4 =  A_U Tr \lambda^2_{U} +
A_D Tr (\lambda^2_{D} + {1 \over 3} \lambda^2_{L}) + 8 \mu Tr (\lambda_U
{\cal A_U} + \lambda_D {\cal A_D} + {1 \over 3} \lambda_L {\cal A_L}) v_1
v_2 \ , \label{eq:ar54} \eal  \eea
where $v_1$ and $v_2$ are the vacuum expectation values of the two
Higgs doublets. In (\ref{eq:ar54}) ${\cal A_U}$, ${\cal A_D}$ and ${\cal
A_L}$ are trilinear soft breaking terms and the trace is in the family
space. $A_U$ and $A_D$ are given by the expressions \bea
\bal A_U = 2 \ [ 2\mu^2 / tg^2 \beta +
4 M^2 - M^2_Z + (g^2_1 + g^2_2) v^2_1 ] \ v^2_2 \ , \\ \\ A_D = 2 \ [ 2\mu^2 \
tg^2 \beta + 4 M^2 - M^2_Z + (g^2_1 + g^2_2) v^2_2 ] \ v^2_1 \ ,
\label{eq:ar55} \eal 
\eea
where  $g_1,g_2$ are the $U(1)$, $SU(2)$ gauge couplings, $M$ is a
universal squark soft mass and $M^2_Z =  {1 \over 2} (g^2_1 + g^2_2 )
(v^2_1 + v^2_2 ) $ is the $Z$  mass. In order to show that $A_U,A_D>0$,
one may use the phenomenological inequality  
\bea
 (Str \cm^2)_{\mbox{quarks + squarks}} = 4M^2 > M^2_Z \ .
\label{eq:ar56}  \eea
The vacuum energy (\ref{eq:ar101}) has roughly
the Nambu form \cite{N} with 
an additional linear term which does not change the shape 
of the vacuum energy as a function of the Yukawas, but which plays an
essential role in the minimization process. 

The positivity of $A_U$, $A_D$ is a consequence of supersymmetry
in the sense that it is due to the Yukawa dependent bosonic contributions
in (\ref{eq:ar54}). In the non-supersymmetric Standard Model the sign
is negative and the present considerations do not apply.
Using eq.(\ref{eq:ar101}) and eq.(\ref{eq:ar102}), we obtain the vacuum
energy as a function of the matrices $\lambda_U$ and $\lambda_D$, which is a
paraboloid unbounded from below. If the minimization is freely performed,
then they are driven to the maximally allowed values and no hierarchy
is generated. 

Consider now the mass matrices (\ref{eq:ar21}) with 
$\lambda \sim (t_1 / t_2)^{1 \over 2}$ a dynamical parameter to be
determined by the minimization.  
We discussed in Ref.\cite{BD} two types of constraints: (a) the
proportionality constraint where one of the couplings is proportional
to another (to some positive power) $\lambda_1 = {\rm cst}\cdot
\lambda_2^n,\; n>0$, (b) a {\em multiplicative constraint} where the
product of two couplings (or positive powers of them) is fixed to be a
moduli independent constant: $\lambda_1 \lambda_2^n = {\rm cst}, \;
n>0$. Only the second constraint leads to dynamical hierarchy of
couplings. Fortunately for $x,y > 0$ in (\ref{eq:ar21}) we get the second
type of constraints, for example $(\hl^{33}_U)^y  (\hl^{33}_D)^x
= cst$. In this case if $\hl^{33}_U$ for example is big, the
constraint (valid at $M_P$) forces $\hl^{33}_D$ to be small and we
naturally obtain small numbers.

For the case of two moduli, the conditions to have $x > 0$, $y > 0$ read
\bea
n_{Q_3} + n_{U_3} + n_{H_2} > -3/2 \ , n_{Q_3} + n_{D_3} + n_{H_1} <
-3/2 \label{eq:ar57} \eea
and they should be fulfilled in order to obtain multiplicative-type constraints.
An interesting case (treated in detail in \cite{BD}, where we keeped only
$\lambda^{33}_U$ and $\hl^{33}_D$ in the computations) is $x = y$.
The relevant constraints are then symmetric in the up and down quarks. 

The low energy effective potential is to be minimized with respect to $\lambda$.
For this the RG equations are used in order to translate the structures 
(\ref{eq:ar21})
from $M_P$ to $\mu$. The analysis is essentially the same as in \cite{BD}, the
whole structure of the mass matrices does not change qualitatively the
results. There are essentially two conditions for the top quark to be
the heaviest fermion. The first is (for $g_1 = 0$) 
\bea
tg^2 \beta > {2 M^2 + m_1^2 \over 2 M^2 + m_2^2 } \ . \label{eq:ar58} 
\eea
where $m_1,m_2$ are the supersymmetric mass terms for the two Higgs.
The second is a rather involved lower bound for the dilaton vacuum
expectation value, so that  the underlying string theory must be in a
perturbative regime. We therefore need a minimal critical value for $tg
\beta$ of order one,  which depends on the soft masses, in order to have
a heavy top quark. Under these two assumptions, there is no need of fine
tuning to obtain a value of $\lambda$ of order $0.2$ which allows to
understand the hierarchy between the top quark and the other fermions.
 \section{Concluding remarks.}

In this paper we analyzed the structure of the fermion mass matrices in
the effective superstring theories. It is found that, in some cases of
phenomenological interest, they are similar to the structures obtained
by imposing abelian horizontal symmetries. The analog of the abelian
charges are the modular weights of the matter fields; the small 
expansion parameters are provided by the vev's of some moduli fields
away from their self-dual values. 
It is known that by imposing a horizontal $U(1)$ symmetry in effective orbifold
models we obtain a relation between modular weights and the $U(1)$ charges of
different families \cite{DPSA}. We found here \cite{BDU} that even without
 a horizontal symmetry,
the fermion mass matrices can present similar structures, if the theory have
a diagonal modular symmetry (\ref{eq:ar10'}).
Hierarchical structures for the mass matrices
are obtained by assigning different modular weights for the three families
of quarks and leptons with respect to some moduli fields. A particular
case of interest is when the Yukawas are homogeneous functions of the
moduli, which can be viewed as a consequence of a 'diagonal' modular
symmetry of the theory, in the case where the original string couplings are
pure numbers. An interesting consequence is that the squark and
slepton mass matrices and the trilinear soft terms are proportional to the
 identity matrix. Consequently
they give no contributions to the FCNC processes like $b \rightarrow s \gamma$
or $\mu \rightarrow e \gamma$.

We stressed an intriguing connection between the mass matrices and the modular
anomalies, similar to the one between mass matrices and
mixed gauge anomalies in the horizontal symmetry approach recently
discussed in the literature. This is probably an additional argument in
favor of a close relationship between horizontal symmetries and modular
symmetries in effective string models.
 A phenomenologically relevant mass spectrum
requires one-loop modular anomalies, which can be cancelled in two ways.
The first one is the Green-Schwarz mechanism of superstrings. In this context,
if the Yukawa couplings are homogeneous functions of moduli and if the sum of the modular
weights of the two Higgs doublets of the MSSM is symmetric in the moduli, then
a correct mass pattern asks for a Green-Schwarz mechanism with $k_1 = {5
\over 3}$ and the Weinberg angle is predicted to be $\sin^2 \theta_W
= {3 \over 8}$. The second way uses the moduli dependent threshold 
corrections to the gauge coupling constants. In this case we obtain a relation
between the fermion masses, modular weights and the unification scale
$M_U$. Our analysis shows that we can acommodate a low value $M_U \sim M_s / 50$
provided the Higgs modular weights satisfy a constraint which is allowed
at Kac-Moody level two or three in abelian orbifolds. Hence we have the
possibility of a succesful unification scheme.

We have also investigated a dynamical mechanism for understanding the
fermion masses as a low-energy minimization process, previously restricted
to the top and bottom couplings. We show that the mechanism is easily 
generalized to account
for the whole structure of the mass matrices, provided two inequalities on
the modular weights hold.

We have given in \cite{BDU} orbifold
examples where the hierarchies of the type that we propose are
allowed. There are no examples at Kac-Moody level one due to the limited 
range of the allowed modular weights, but we give examples at level two and 
three.

There are, of course, many open questions and problems which deserve further
investigations. First of all the vev's of the moduli fields should be fixed
by the dynamics, which usually prefers the self-dual points. In the dynamical
approach, it would be also interesting to view the determination of the
Yukawa couplings directly from the point of view of the moduli fields:
in particular why the corresponding flat directions remain unlifted down to
low energies.

Finally it would be interesting to construct explicit orbifold models with 
hierarchical mass matrices along these lines and to investigate their
phenomenological virtues. 
  
\vskip 1.2cm
{\bf Acknowledgements}
\vskip .8cm
It is a pleasure to thank P. Bin\'etruy for his collaboration which led to the
present work and S. Pokorski and C.A. Savoy for many interesting discussions and
comments.

\end{document}